\documentclass[reprint,amsmath,amssymb,pra,superscriptaddress]{revtex4-1}
\usepackage{graphicx}
\usepackage{bm}
\usepackage{booktabs}

\begin{document}
\title{Critical Dynamics of an Asymmetrically\\Bidirectionally Pumped Optical Microresonator}

\author{Jonathan~M.~Silver}
 \email{jonathan.silver@npl.co.uk}
\affiliation{National Physical Laboratory, Hampton Road, Teddington TW11 0LW, UK}
\affiliation{City, University of London, Northampton Square, London EC1V 0HB, UK}
\author{Kenneth~T.~V.~Grattan}
\affiliation{City, University of London, Northampton Square, London EC1V 0HB, UK}
\author{Pascal~Del'Haye}
\affiliation{National Physical Laboratory, Hampton Road, Teddington TW11 0LW, UK}

\date{\today}

\begin{abstract}
An optical ring resonator with third-order, or Kerr, nonlinearity will exhibit symmetry breaking between the two counterpropagating circulating powers when pumped with sufficient power in both the clockwise and counterclockwise directions. This is due to the effects of self- and cross-phase modulation on the resonance frequencies in the two directions. The critical point of this symmetry breaking exhibits universal behaviors including divergent responsivity to external perturbations, critical slowing down, and scaling invariance. Here we derive a model for the critical dynamics of this system, first for a symmetrically-pumped resonator and then for the general case of asymmetric pumping conditions and self- and cross-phase modulation coefficients. This theory not only provides a detailed understanding of the dynamical response of critical-point-enhanced optical gyroscopes and near-field sensors, but is also applicable to nonlinear critical points in a wide range of systems.
\end{abstract}

\maketitle
\section{Introduction}
Spontaneous symmetry breaking is ubiquitous in physics, occurring at every possible energy scale all the way from the Higgs mechanism~\cite{higgs1964broken} down to superconductivity~\cite{PhysRev.108.1175}, superfluidity~\cite{landau1941theory} and other exotic quantum states of matter at ultracold temperatures~\cite{Bloch2008}. Associated with every spontaneous symmetry-breaking transition is a critical point~\cite{stanley1971phase} -- a point in parameter space on the boundary of the symmetry-broken regime where the symmetric state of the system is neither stable nor unstable, which exhibits certain universal features. Firstly, the system will have divergent responsivity to external perturbations that break the symmetry of the system, exhibiting large excursions in response to tiny perturbations but always eventually returning to the symmetric state if the perturbation is removed. Secondly, the characteristic timescales and lengthscales (where relevant) of the system's response diverge -- the system exhibits fluctuations at all lengthscales that decay with time according to a power law rather than exponentially and thus take exponentially longer to reach the steady state. This is often referred to as critical slowing down~\cite{stanley1971phase}. Thirdly, the equations of the system around the critical point exhibit scaling invariance, meaning that they are unchanged when the offsets of the various parameters from the critical point, as well as length and time, are scaled by certain powers of each other.

One system that exhibits spontaneous symmetry breaking of the discrete group $Z_2$ is that of a symmetrically bidirectionally-pumped optical microresonator with Kerr nonlinearity, in which the counterpropagating circulating powers will spontaneously deviate from each other~\cite{del2017symmetry, Cao2017, Woodley2018, DelBino2018}. This is an example of what is known as a pitchfork bifurcation due to the way in which one stable solution splits into two. Because the symmetry in question is clockwise-counterclockwise symmetry, the system, at a critical point, exhibits divergent responsivity to perturbations that distinguish between the two directions, including most notably pump power and detuning differences. Since pump detuning differences can be induced via the Sagnac effect by rotating the entire setup, which causes the counterpropagating resonance frequencies to differ by an amount proportional to the rotation velocity~\cite{Post1967}, this critical point can be used to create a simple yet extremely sensitive gyroscope~\cite{Kaplan1981, Wang2014a, silver2019critical}. The Sagnac effect is related to but distinct from Fizeau drag, which was recently demonstrated in a similar experiment in which the microresonator was rotated but the rest of the setup remained stationary~\cite{Maayani2018}.

The development of such a gyroscope, or indeed other critical-point-enhanced sensors such as for refractive index~\cite{Wang2015,svela2019spontaneous} requires a detailed understanding of the critical dynamics of the system, including its response to time-dependent and finite-amplitude inputs. In this paper we show that in the limit as we approach the critical point, the dynamics are governed by a simple equation, from which the divergent responsivity, critical slowing down, and scaling invariance are manifestly apparent. This is done first in Section~\ref{sec:symm} for the simplest case of a symmetrically-pumped resonator with a Kerr cross-phase modulation (XPM) coefficient twice as large as that of self-phase modulation (SPM), as is the case in any dielectric solid~\cite{Boyd}. In Section~\ref{sec:asymm} we show that the same critical point and behavior can occur even when the system itself is not symmetric, but when two different asymmetries, for example in pump power and detuning, balance each other~\cite{garbin2019asymmetric}. We derive the exact condition for the critical point, as well as the equation for the critical dynamics, in an asymetrically-pumped resonator with arbitrary and even asymmetric SPM and XPM coefficients.

The theory presented here applies not only to Kerr-related symmetry breaking between counterpropagating light, but also between different frequencies, propagation angles~\cite{Haelterman1991}, and opposite circular polarisations~\cite{areshev1983polarization, haelterman1994polarization, Copie2019}, all of which obey the same equations. For instance, the asymmetric critical point was recently demonstrated for the polarisation case in a fiber loop cavity~\cite{garbin2019asymmetric}. Furthermore, this theory applies to systems where the Kerr effect is substituted with a Kerr-like interaction such as the magnetic nonlinearity~\cite{Martin2010}, or even to similar nonlinear systems outside the optical domain altogether.

The ratio between the XPM and SPM coefficients can take different values in different materials, including less than two in semiconductors and gases due to diffusive effects, and as much as seven for interaction between opposite circular polarisations in Kerr liquids~\cite{Boyd, hill2019effects}. Differences between the two mode volumes will lead to asymmetries in both the SPM and XPM coefficients, while asymmetric effective SPM coefficients but symmetric XPM coefficients can arise if the light in one of the modes is not monochromatic~\cite{silver2019critical}.

Finally, a condition is derived for decoupling the critical dynamics from the thermal nonlinearity~\cite{Carmon2004}, which although perfectly symmetric in its action, is typically much larger than the Kerr effect, and could thus disrupt the critical dynamics in the case of asymmetric pumping conditions or SPM or XPM coefficients.

\section{Symmetric pumping conditions}\label{sec:symm}
When an optical ring resonator with Kerr, or $\chi^{(3)}$, nonlinearity is pumped with light of equal power and frequency in both directions, a spontaneous splitting can occur between the two counterpropagating circulating powers and resonance frequencies~\cite{del2017symmetry, Cao2017, Woodley2018}. This occurs due to the interplay between the circulating-power-dependent Kerr shifts of the counterpropagating resonance frequencies and the detuning-dependent circulating powers due to the pump frequency being on the side of the resonance. The Kerr effect decreases each resonance frequency by an amount proportional to the circulating power in that mode (from SPM) plus twice that in the counterpropagating mode (from XPM). This means that the resonance frequency is lower in the direction with less circulating power. If the pump is blue-detuned from the resonance -- a necessary condition for passive thermal locking of the resonance to the pump frequency~\cite{Carmon2004} -- then the direction with less circulating power will be shifted further from the pump, which in turn increases the circulating power difference, creating positive feedback that causes the symmetry to spontaneously break.

\begin{figure}
\includegraphics*[width=0.95\columnwidth]{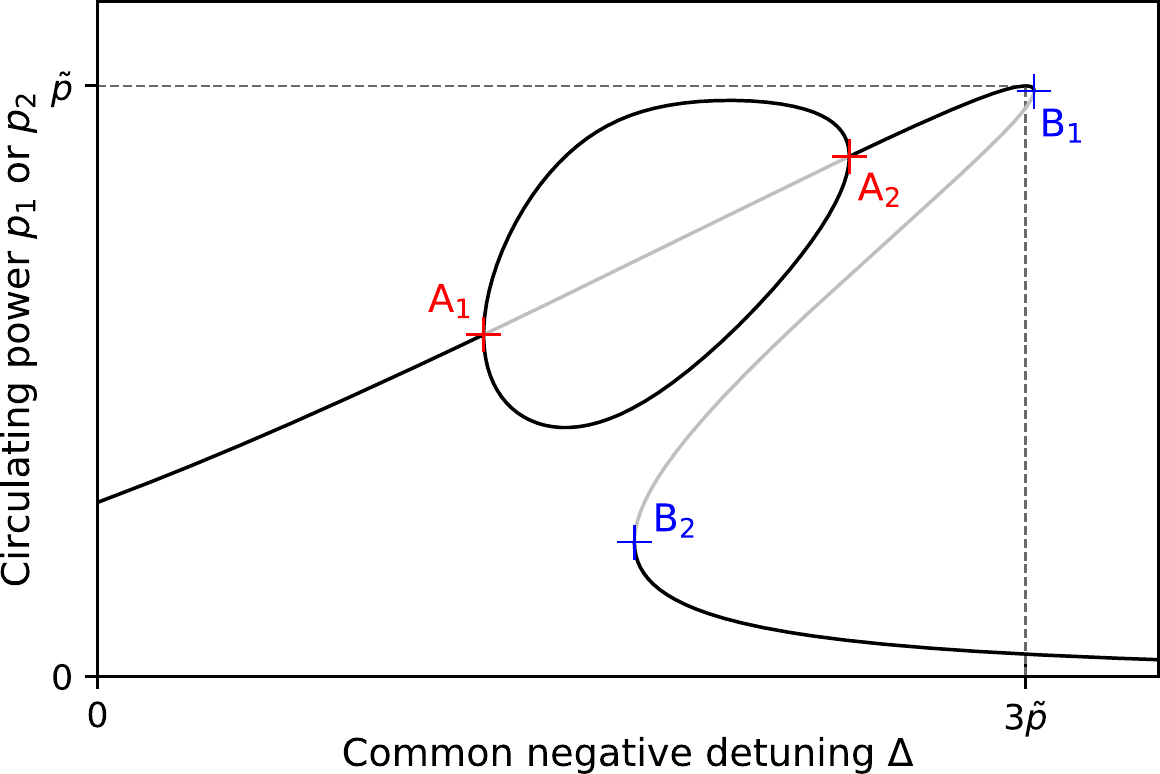}
\caption{Solutions to Eq.~(\ref{eq:ssLorentz}) under symmetric pumping conditions $\tilde{p}_{1,2}=\tilde{p}$ and $\Delta_{1,2}=\Delta$, illustrated for $\tilde{p}=1.75$. Solid black and grey lines represent stable and unstable solutions respectively. Between the critical points $\text{A}_1$ and $\text{A}_2$, the symmetric solution $p_1=p_2$ is unstable and two new symmetry-broken stable solutions appear in which $p_1$ and $p_2$ take the opposite branches shown. Another unstable region lies between $\text{B}_1$ and $\text{B}_2$, which occurs even for a unidirectionally-pumped resonator and here corresponds to a symmetric bistability.\label{fig:bubble}}
\end{figure}

This effect may be described by solving the following pair of simultaneous equations for the circulating powers $p_{1,2}$ in the two counterpropagating directions in terms of the pump powers $\tilde{p}_{1,2}$ and detunings $\Delta_{1,2}$ from the resonance without Kerr shift~\cite{Woodley2018}:
\begin{equation}
p_{1,2}=\frac{\tilde{p}_{1,2}}{1+(p_{1,2}+2p_{2,1}-\Delta_{1,2})^2}.\label{eq:ssLorentz}
\end{equation}
Here, and throughout this paper, we use the dimensionless quantities defined in Table~\ref{tab:dlqs}. Equation~(\ref{eq:ssLorentz}) is simply the dimensionless form of the Lorentzian resonance curves for the circulating powers, taking into account the Kerr shifts. Note the factor of two in front of the counterpropagating circulating power, corresponding to the ratio betweeen the strengths of XPM and SPM in a dielectric solid with Kerr nonlinearity; this ratio is generalised in Section~\ref{sec:asymm}.
\begin{table}
\caption{Dimensionless quantities used in this manuscript. $\eta_\text{in}$ is the resonant in-coupling efficiency equal to $4\kappa\gamma_0/\gamma^2$ where $\kappa$, $\gamma_0$ and $\gamma=\gamma_0+\kappa$ are the coupling, intrinsic and total half-linewidths respectively. $P_{\text{in,}1,2}$ and $P_{\text{circ,}1,2}$ are the pump and circulating powers respectively. $P_0 = \pi n_0^2V/(n_2\lambda QQ_0)$ is the characteristic in-coupled power required for Kerr nonlinear effects, where $n_0$ and $n_2$ are the linear and nonlinear refractive indices, $V$ is the mode volume, and $Q=\omega_0/(2\gamma)$ and $Q_0=\omega_0/(2\gamma_0)$ are the loaded and intrinsic quality factors respectively for cavity resonance frequency $\omega_0$ (without Kerr shift). $\mathcal{F}_0 = \Delta\omega_\text{FSR}/(2\gamma_0)$ is the cavity's intrinsic finesse for free spectral range $\Delta\omega_\text{FSR}$, and $\omega_{1,2}$ are the pump frequencies.}
\label{tab:dlqs}
  \begin{center}
    \begin{tabular}{clc}
      \toprule
      \textbf{Symbol} & \textbf{Description} & \textbf{Formula}\\
      \midrule
      $\tilde{p}_{1,2}$ & Pump powers & $\eta_\text{in}P_{\text{in,}1,2}/P_0$\\
      $p_{1,2}$ & Circulating powers & $2\pi P_{\text{circ,}1,2}/(\mathcal{F}_0 P_0)$\\
      $\Delta_{1,2}$ & \parbox{4cm}{\raggedright \vspace{1.2mm} Pump detunings from resonance frequency without Kerr shift\strut} & $(\omega_0-\omega_{1,2})/\gamma$\\
      $\tilde{e}_{1,2}$ & Pump field amplitudes & $\tilde{p}_{1,2} = \left|\tilde{e}_{1,2}\right|^2$\\
      $e_{1,2}$ & Circulating field amplitudes & $p_{1,2} = \left|e_{1,2}\right|^2$\\
      \bottomrule
    \end{tabular}
  \end{center}
\end{table}
Under symmetrical pumping conditions $\tilde{p}_{1,2}=\tilde{p}$ and $\Delta_{1,2}=\Delta$, symmetry breaking occurs for a range of $\Delta$ if $\tilde{p}$ exceeds $8/(3\sqrt{3})\simeq1.54$~\cite{del2017symmetry, Woodley2018}. This is illustrated in Fig.~\ref{fig:bubble} for $\tilde{p}$ a little above this threshold. As the detuning approaches the symmetry-broken regime, the difference between $p_1$ and $p_2$ exhibits increasing responsivity to perturbations that break the directional symmetry, such as pump power or detuning differences. This responsivity diverges at each of the critical points $\text{A}_1$ and $\text{A}_2$, at which the finite-amplitude response is proportional to the cube root of the perturbation.

We begin with the dimensionless equations for the time derivatives $\dot{e}_{1,2}$ of the electric field amplitudes $e_{1,2}$ in the two counterpropagating modes in the rotating frames of their respective pump fields~\cite{Woodley2018}, which yield Eq.~({\ref{eq:ssLorentz}) in the steady state:
\begin{equation}
\dot{e}_{1,2} = \tilde{e}_{1,2} - \left(1+i\left( |e_{1,2}|^2 + 2|e_{2,1}|^2 - \Delta_{1,2}\right)\right)e_{1,2}.\label{eq:dynEq}
\end{equation}
Once again, the notation is given in Table~\ref{tab:dlqs}, and time is in units of $1/\gamma$. We shall assume for now that $\tilde{e}_1 = \tilde{e}_2 = \tilde{e}$ is constant with time, while $\Delta_{1,2}$ undergo small time-dependent perturbations around a common value $\Delta$, with $\tilde{e}$ and $\Delta$ chosen so as to place the system at a critical point. We are interested in the perturbative dynamics of $e_{1,2}$ around a symmetric steady-state solution \mbox{$e_{1,2}$ = $e$} that satisfies $\tilde{e}=(1+i(3p-\Delta))e$, where the circulating power in each direction $p=|e|^2$. Choosing the phase of $\tilde{e}$ such that $e$ is real and positive, we let
\begin{equation}\begin{aligned}
\Delta_{1,2} &= \Delta + \delta_{1,2}\\
e_{1,2} &= e + f_{1,2} + i g_{1,2},
\end{aligned}\end{equation}
where $\delta_{1,2}$, $f_{1,2}$ and $g_{1,2}$ are all real, and represent perturbations to the pump detunings and circulating field amplitudes and phases respectively. Substituting these into Eq.~(\ref{eq:dynEq}), we can express the dynamical equations of $f_{1,2}$ and $g_{1,2}$ in the form
\begin{equation}
\dot{\mathbf{f}} = \mathbf{M f} + \mathbf{d} + \mathbf{D f} + \mathbf{k} + \mathbf{l},\label{eq:vecDynamics}
\end{equation}
where
\begin{align}
\mathbf{f} &= \begin{pmatrix}f_1\\g_1\\f_2\\g_2\end{pmatrix}\!;\quad\mathbf{d} = e\begin{pmatrix}0\\\delta_1\\0\\\delta_2\end{pmatrix}\!;\quad
\mathbf{D} = \begin{pmatrix}0&-\delta_1&0&0\\\delta_1&0&0&0\\0&0&0&-\delta_2\\0&0&\delta_2&0\end{pmatrix}\nonumber\\
\mathbf{M} &= \begin{pmatrix}-1&3p-\Delta&0&0\\\Delta-5p&-1&-4p&0\\0&0&-1&3p-\Delta\\-4p&0&\Delta-5p&-1\end{pmatrix}\nonumber\\
\mathbf{k} &= e\begin{pmatrix}g_1\left(2f_1+4f_2\right)\\-\left(3f_1^2+g_1^2+2f_2^2+2g_2^2+4f_1f_2\right)\\g_2\left(2f_2+4f_1\right)\\-\left(3f_2^2+g_2^2+2f_1^2+2g_1^2+4f_1f_2\right)\end{pmatrix}\nonumber\\
\mathbf{l} &= \begin{pmatrix}g_1\left(f_1^2+g_1^2+2f_2^2+2g_2^2\right)\\-f_1\left(f_1^2+g_1^2+2f_2^2+2g_2^2\right)\\ g_2\left(f_2^2+g_2^2+2f_1^2+2g_1^2\right)\\-f_2\left(f_2^2+g_2^2+2f_1^2+2g_1^2\right)\end{pmatrix}\!.
\end{align}

We begin by considering the linear response of the system around the steady-state solution, which is governed by $\dot{\mathbf{f}} = \mathbf{M f} + \mathbf{d}$. Here we have kept only the terms that are first-order in the perturbations $f_i$ , $g_i$ and $\delta_i$, discarding those that are second- or third-order. Inspecting the eigenvalues of $\mathbf{M}$, we find that the steady-state solution is unstable when one of the following two conditions is satisfied, as each condition causes a different eigenvalue to be real and positive:
\begin{align}
(p-\Delta)(3p-\Delta)&<-1\label{eq:sbinstab}\\
(3p-\Delta)(9p-\Delta)&<-1\label{eq:symminstab}.
\end{align}
Since $p>0$, Eq.~(\ref{eq:sbinstab}) can hold only when \mbox{$3p-\Delta>0$}, and (\ref{eq:symminstab}) only when \mbox{$3p-\Delta<0$}. Since $3p-\Delta$ is the laser detuning from the Kerr-shifted resonance, this means that (\ref{eq:sbinstab}) must correspond to the symmetry-broken region between the critical points $\text{A}_1$ and $\text{A}_2$ in Fig.~\ref{fig:bubble}, and (\ref{eq:symminstab}) to the region between $\text{B}_1$ and $\text{B}_2$. The critical points are thus characterised by the boundary of~(\ref{eq:sbinstab}):
\begin{equation}
(p-\Delta)(3p-\Delta)=-1.\label{eq:critPtCond}
\end{equation}
Under this condition, which shall be assumed to hold for the rest of this section, the eigenvectors $\mathbf{v}_i$ and corresponding eigenvalues $\lambda_i$ of $\mathbf{M}$ are:
\begin{equation}\begin{aligned}
\mathbf{v}_1&=\begin{pmatrix}-a\\-1\\a\\1\end{pmatrix}\!,\quad\lambda_1 = 0;\;\quad\mathbf{v}_2=\begin{pmatrix}a\\-1\\-a\\1\end{pmatrix}\!,\quad\lambda_2 = -2;\\
\mathbf{v}_3&=\begin{pmatrix}-ir\\1\\-ir\\1\end{pmatrix}\!,\quad\lambda_3 = -1+ia/r;\\
\mathbf{v}_4&=\begin{pmatrix}ir\\1\\ir\\1\end{pmatrix}\!,\quad\lambda_4 = -1-ia/r.\label{eq:eigens}
\end{aligned}\end{equation}
where $a=3p-\Delta$ and $r=\sqrt{(3p-\Delta)/(9p-\Delta)}$ are real and positive. The slow critical dynamics will thus be dominated by $\mathbf{v}_1$ as this has a zero eigenvalue, whereas the other three have eigenvalues with negative real parts of order unity, and thus decay away on a timescale of the order of the cavity lifetime. Note that $\mathbf{v}_1$ corresponds to an antisymmetric combined amplitude and phase perturbation.

Turning again to Eq.~(\ref{eq:vecDynamics}) including all its nonlinear terms, we will now express it in this eigenbasis by using the inverse basis $\{\mathbf{u}_i\}: \mathbf{u}_i\cdot\mathbf{v}_j = \delta_{ij}$, where $\delta_{ij}$ is the Kronecker delta, to decompose 
\begin{equation}\begin{aligned}
&\mathbf{f} = \sum_i\mu_i\mathbf{v}_i,\quad \mathbf{d} = \sum_id_i\mathbf{v}_i,\\ &\mathbf{D} = \sum_{i,j}D_{ij}\mathbf{v}_i\mathbf{u}_j^\text{T},\quad \mathbf{k} = \sum_{i,j,k} K_{ijk}\mathbf{v}_i\mu_j\mu_k,\\ &\text{and}\quad\mathbf{l} = \sum_{i,j,k,l} L_{ijkl}\mathbf{v}_i\mu_j\mu_k\mu_l,
\end{aligned}\end{equation}
 where $i,j,k,l$ index the eigenvectors and hence run from 1 to 4, and $\mu_i$ is the projection of $\mathbf{f}$ along $\mathbf{v}_i$:
\begin{equation}\begin{aligned}
\dot{\mu}_i\;=\;\,& \lambda_i\mu_i + d_i + \sum_jD_{ij}\mu_j\\&+ \sum_{j,k}K_{ijk}\mu_j\mu_k + \sum_{j,k,l}L_{ijkl}\mu_j\mu_k\mu_l.\label{eq:eigenDynamics}
\end{aligned}\end{equation}
To extract the dynamics in the region immediately surrounding the critical point, we will start by removing the driving terms $d_i$ and $D_{ij}$:
\begin{equation}
\dot{\mu}_i\;=\;\lambda_i\mu_i + \sum_{j,k}K_{ijk}\mu_j\mu_k + \sum_{j,k,l}L_{ijkl}\mu_j\mu_k\mu_l.\label{eq:eigenDynNoDriv}
\end{equation}
For small perturbations and responses around the critical point, we may say that $|\mu_1|\ll1$. Furthermore, since $\mu_2$, $\mu_3$, and $\mu_4$, unlike $\mu_1$, have exponential decay times that are short compared to the timescale of the critical dynamics as discussed above, it is safe to assume that $|\mu_i|\ll|\mu_1|,\;i\ne1$. Nevertheless, we shall see that these cannot be ignored entirely as they still contribute to the dynamics of $\mu_1$. Looking at the case $i=1$ in Eq.~(\ref{eq:eigenDynNoDriv}), since $\lambda_1 = 0$, the leading term in $\dot{\mu}_1$ would be $K_{111}\mu_1^2$, however $K_{111}=0$ by considerations of directional symmetry i.e.\ switching the 1 and 2 directions. This leaves
\begin{equation}
\dot{\mu}_1\;=\;2\sum_{i\ne1} K_{11i}\mu_1\mu_i + L_{1111}\mu_1^3\label{eq:mu1dotLO}
\end{equation}
to leading order, assuming that $K_{ijk}=K_{ikj}$ by construction. Looking again at Eq.~(\ref{eq:eigenDynNoDriv}), we can see that to leading order, the other $\mu_i$ obey the following quasi-steady-state equations:
\begin{equation}
0=\lambda_i\mu_i+K_{i11}\mu_1^2,\quad i\ne1.
\end{equation}
Noting that also $K_{211}=0$ by directional symmetry, this can be combined with Eq.~(\ref{eq:mu1dotLO}) to give
\begin{equation}
\dot{\mu}_1\;=\;\left(L_{1111}-2\sum_{i=3,4} \frac{K_{11i}K_{i11}}{\lambda_i}\right)\mu_1^3.\label{eq:mu1dotEffLO}
\end{equation}
Both terms are of equal order, so indeed we cannot neglect the effect of $\mu_3$ and $\mu_4$ on the dynamics of $\mu_1$. Futhermore, we may observe that $\mu_{3,4}$ scale as $\mu_1^2$ and $\dot{\mu}_1$ scales as $\mu_1^3$, the latter confirming that the timescale of the dynamics increases (as $\mu_1^{-2}$) as we zoom closer and closer into the critical point.

Equation~(\ref{eq:mu1dotEffLO}) describes the free evolution of the system to leading order. Now we re-introduce the driving terms $d_i$ and $D_{ij}$ at magnitudes that preserve the above hierarchy of scalings. For this, it is useful to introduce the common- and differential-mode detunings $\delta_\text{c} = (\delta_1+\delta_2)/2$ and $\delta_\text{d} = (\delta_1-\delta_2)/2$, and to note that
\begin{equation}
d_1 = -\frac{e\delta_\text{d}}{2},\quad d_{3,4} = \frac{e\delta_\text{c}}{2},\quad D_{11} = (2p-\Delta)\delta_\text{c}.
\end{equation}
Substituting the first two of these into Eq.~(\ref{eq:eigenDynamics}) for the relevant $i$, we deduce that $\delta_\text{d}$ scales as $\mu_1^3$ and $\delta_\text{c}$ as $\mu_1^2$, and therefore that the only element of $ D_{ij}$ that can possibly affect the dynamics to leading order is $D_{11}$. This leaves us, to leading order, with
\begin{equation}
\mu_i = -\frac{K_{i11}\mu_1^2+ e\delta_\text{c}/2}{\lambda_i}\quad\text{for}\quad i=3,4,\label{eq:mu34}
\end{equation}
satisfying $\mu_4 = \mu_3^*$ since $K_{114} = K_{113}^*$ and $\lambda_4 = \lambda_3^*$, and
\begin{align}
\dot{\mu}_1&=-\frac{e\delta_\text{d}}{2}+\left(\!(2p-\Delta)-e\!\left(\!\frac{K_{113}}{\lambda_3}+\frac{K_{114}}{\lambda_4}\!\right)\!\right)\!\delta_\text{c}\mu_1\nonumber\\&\;\;\;+\left(\!L_{1111}-2\!\left(\!\frac{K_{113}K_{311}}{\lambda_3}+\frac{K_{114}K_{411}}{\lambda_4}\!\right)\!\right)\!\mu_1^3\label{eq:mu1dotfinal}\\[10pt] &\;=-\frac{e\delta_\text{d}}{2}\!+\!\frac{5p\!-\!2\Delta}{4}\delta_\text{c}\mu_1\!+\!\frac{\left(3p\!-\!\Delta\right)\!\left(4\!+\!4p\Delta\!-\!15p^2\right)}{2}\mu_1^3.\nonumber
\end{align}
So far, for conciseness, we have not considered the effect of pump power purturbations. It turns out that these have a very similar effect to detuning purturbations; their treatment is summarised as follows. We may represent small fractional pump power perturbations $\epsilon_{1,2}$ by letting $\tilde{e}_{1,2}=\tilde{e}\left(1+\epsilon_{1,2}/2\right)$ and consequently adding $e\left(\epsilon_1,a\epsilon_1,\epsilon_2,a\epsilon_2\right)\!/2$ to $\mathbf{d}$. Decomposing these into common- and differential-mode components $\epsilon_\text{c}=(\epsilon_1+\epsilon_2)/2$ and $\epsilon_\text{d}=(\epsilon_1-\epsilon_2)/2$ and revisiting the above steps, we find that $\epsilon_{\text{d}}$ scales as $\mu_1^3$ and $\epsilon_{\text{c}}$ as $\mu_1^2$ just as with detuning perturbations, and that Eq.~(\ref{eq:mu1dotfinal}) becomes
\begin{align}
\dot{\mu}_1&=-\frac{e}{2}\left(\delta_\text{d}+p\epsilon_\text{d}\right)+\left(\frac{5p-2\Delta}{4}\delta_\text{c}+\frac{2\Delta-3p}{4}p\epsilon_\text{c}\!\right)\!\mu_1\nonumber\\&\;\;\;\;\;\;\;+\frac{\left(3p-\Delta\right)\!\left(4+4p\Delta-15p^2\right)}{2}\mu_1^3.\label{eq:mu1dotfinalWithPowerPert}
\end{align}
Interestingly, if we include pump \textit{phase} perturbations, for example by allowing $\epsilon_{1,2}$ to be complex, we find that they (as distinct from detuning perturbations which are analogous to their time derivatives) play no role in the critical dynamics to leading order. This is actually expected, since Eq.~(\ref{eq:dynEq}) is invariant under static phase rotations of $\tilde{e}_{1,2}$, as long as the same rotations are applied respectively to $e_{1,2}$.

Importantly, the coefficient of $\mu_1^3$ in Eqs.~(\ref{eq:mu1dotfinal}) and (\ref{eq:mu1dotfinalWithPowerPert}) is always negative, which can be seen by substituting in Eq.~(\ref{eq:critPtCond}) to give $-(3p-\Delta)((2\Delta-5p)^2+2p^2)/2$. We can therefore re-express~(\ref{eq:mu1dotfinalWithPowerPert}) in the form
\begin{equation}
\dot{y}=-y^3+xy+z,\label{eq:effCritDynEq}
\end{equation}
where
\begin{equation}\begin{aligned}
x&=\frac{5p-2\Delta}{4}\delta_\text{c}+\frac{2\Delta-3p}{4}p\epsilon_\text{c}\\
y&=-\sqrt{\frac{\left(3p-\Delta\right)\left(15p^2-4p\Delta-4\right)}{2}}\mu_1\\
z&=\sqrt{\frac{p\left(3p-\Delta\right)\left(15p^2-4p\Delta-4\right)}{8}}\left(\delta_\text{d}+p\epsilon_\text{d}\right).
\end{aligned}\end{equation}
From the expression for $\mathbf{v}_1$ in Eq.~(\ref{eq:eigens}) we can relate $y$ to the observable differential-mode (normalised) coupled power $p_\text{d} = \left(p_1-p_2\right)/2$ to leading order as follows:
\begin{equation}
p_\text{d}\;=\;e\left(f_1-f_2\right)\;=\;\sqrt{\!\frac{8p\left(3p-\Delta\right)}{15p^2-4p\Delta-4}}\,y.
\end{equation}
Observe that Eq.~(\ref{eq:effCritDynEq}) is invariant under the transformation $y\rightarrow\kappa y,\;x\rightarrow\kappa^2 x,\;z\rightarrow\kappa^3 z,\; t\rightarrow\kappa^{-2} t$ where $t$ is time and $\kappa$ is an arbitrary scaling parameter. This scaling invariance, in which the equations look the same when each of the variables is scaled by some power of a common parameter, is a universal feature of critical points in many areas of physics, such as ferromagnetism, superconductivity and liquid-gas transitions~\cite{stanley1971phase}.

\begin{figure}
\includegraphics*[width=0.95\columnwidth]{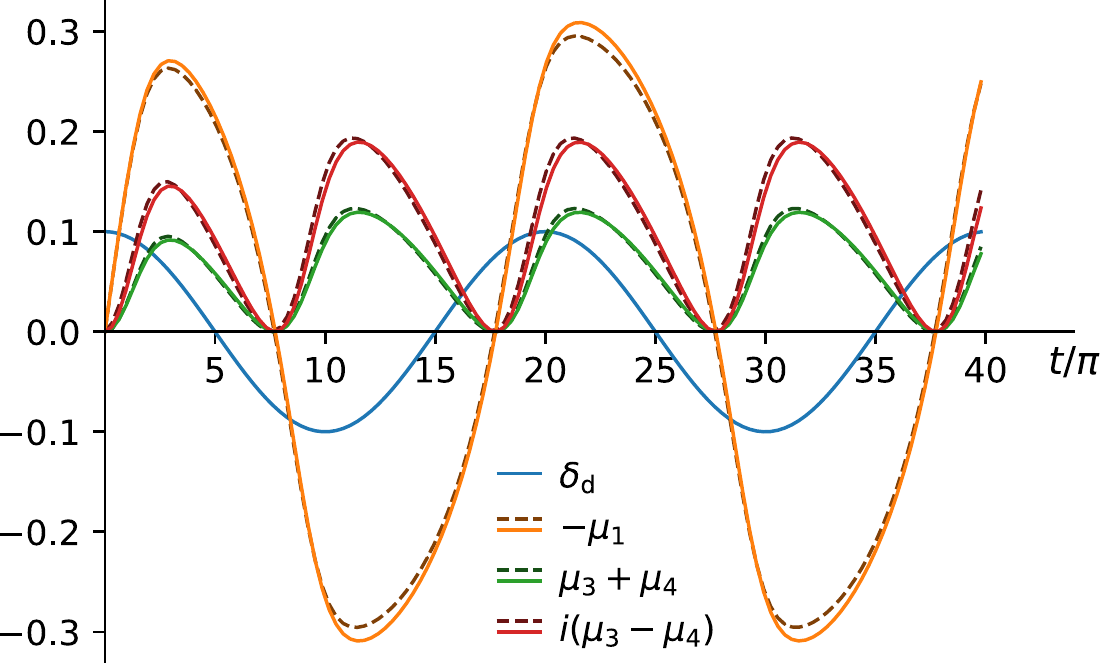}
\caption{Response to a sinusoidally modulated differential-mode detuning $\delta_\text{d}=\delta_\text{d}^\text{AC}\cos{\omega t}$, with $\delta_\text{c} = 0$, at the critical point $p=1$, $\Delta = 2$ for $\delta_\text{d}^\text{AC} = \omega = 0.1$ and $\mu_i = 0$ at $t=0$. Dashed lines indicate exact solutions to Eq.~(\ref{eq:vecDynamics}) resolved into the eigenbasis via $\mu_i = \mathbf{u}_i\cdot\mathbf{f}$, while solid lines are calculated using the leading-order approximation developed in this section, via Eqs.~(\ref{eq:mu34}) and (\ref{eq:mu1dotfinal}). There is a visible difference between the two here since we are not in the regime $|\mu_1|\ll1$, meaning that higher-order terms are not negligible. The smaller $\delta_\text{d}$ and $\mu_1$ are, the better the leading-order approximation becomes, and the smaller $\mu_{3,4}$ become relative to $\mu_1$.\label{fig:dyn}}
\end{figure}

The dynamics of $y$ under Eq.~(\ref{eq:effCritDynEq}) can be summarised as an interplay between three simple behaviors, each of which occurs in its pure form when two of the three terms containing $y$ can be neglected -- cube root ($y=z^{1/3}$), proportional ($y=-z/x$) and integrator ($\dot{y}=z$). Furthermore, Eq.~(\ref{eq:effCritDynEq}) indicates the presence of two universal critical behaviors, namely divergent steady-state responsivity ($|y/z|\rightarrow\infty$ as $|x|, |y|, |\dot{y}|\rightarrow0$) and critical slowing down ($|\dot{y}/y|\rightarrow0$ as $|x|, |y|, |z|\rightarrow0$). Figure~\ref{fig:dyn} shows the dynamics for sinusoidally varying $\delta_\text{d}$, or $z$. Note that the phase lag between $\delta_\text{d}$ and $-\mu_1$ tells us that the system's response time, rather than being around the normalised inverse cavity half-linewidth of unity, is not much less than $1/\omega=10$, due to critical slowing down.

\section{Asymmetric SPM and XPM coefficients and pumping conditions}\label{sec:asymm}
In this section we generalise the theory to asymmetric SPM and XPM coefficients and pump powers and detunings. This is based on an extension of Eq.~(\ref{eq:dynEq}) to general SPM and XPM coefficients $A_{ij}$:
\begin{equation}
\dot{e_j} = \tilde{e}_j-\left(1+i\!\left(\sum_k A_{jk}|e_k|^2-\Delta_j\right)\!\right)e_j,
\end{equation}
where $A_{11}=A_{22}=1,\;A_{12}=A_{21}=2$ reproduces Eq.~(\ref{eq:dynEq}). This time, we expand this around a general asymmetric steady-state solution \mbox{$\Delta_i = \Delta^0_i,\; e_i = e^0_i\in\mathbb{R}^+$,} $\tilde{e}_i = \tilde{e}^0_j = (1+ia_j)e^0_j$ where $a_i=\sum_jA_{ij}p^0_j-\Delta^0_i$, $p^0_i = {e^0_i}^2$. For completeness, in addition to detuning perturbations $\delta_i$, we include fractional pump power and pump phase perturbations, $\epsilon_i$ and $\phi_i$ respectively, from the start (although static phase perturbations $\phi_i$ will again be found to have no effect on the critical dynamics):
\begin{equation}\begin{aligned}
\Delta_i &= \Delta^0_i + \delta_i\\\tilde{e}_j&=\tilde{e}^0_j\left(1+\frac{\epsilon_j}{2}+i\phi_j\right)\!\\ e_j &= e^0_j + f_j + i g_j.
\end{aligned}\end{equation}
We still express the time evolution of $f_i$ and $g_i$ in the form given in Eq.~(\ref{eq:vecDynamics}), but with the following modifications:
\begin{align}
\mathbf{d} &= \!\begin{pmatrix}e^0_1\varepsilon_1\\e^0_1(\delta_1+\zeta_1)\\e^0_2\varepsilon_2\\e^0_2(\delta_2+\zeta_2)\end{pmatrix}\!,\quad\mathbf{M} = \!\begin{pmatrix}-1&a_1&0&0\\-b_1&-1&-c_1&0\\0&0&-1&a_2\\-c_2&0&-b_2&-1\end{pmatrix}\!,\nonumber\\
\mathbf{k} &= \!\begin{pmatrix}2g_1(A_{11}\,e^0_1f_1+A_{12}\,e^0_2f_2)\\-e^0_1(A_{11}(3f_1^2\!+\!g_1^2)\!+\!A_{12}(f_2^2\!+\!g_2^2))\!-\!2A_{12}\,e^0_2f_1f_2\\2g_2(A_{22}\,e^0_2f_2+A_{21}\,e^0_1f_1)\\-e^0_2(A_{22}(3f_2^2\!+\!g_2^2)\!+\!A_{21}(f_1^2\!+\!g_1^2))\!-\!2A_{21}\,e^0_1f_1f_2\end{pmatrix}\!,\nonumber\\
\mathbf{l} &= \!\begin{pmatrix}g_1(A_{11}(f_1^2+g_1^2)+A_{12}(f_2^2+g_2^2))\\-f_1(A_{11}(f_1^2+g_1^2)+A_{12}(f_2^2+g_2^2))\\ g_2(A_{21}(f_1^2+g_1^2)+A_{22}(f_2^2+g_2^2))\\-f_2(A_{21}(f_1^2+g_1^2)+A_{22}(f_2^2+g_2^2))\end{pmatrix}\!,
\end{align}
where $\varepsilon_i=\epsilon_i/2-a_i\phi_i$, $\zeta_i=a_i\epsilon_i/2+\phi_i$, $b_i=a_i+2A_{ii}p^0_i$, $c_1=2A_{12}e^0_1e^0_2$ and $c_2=2A_{21}e^0_1e^0_2$.
The condition for one of the eigenvalues of $\mathbf{M}$ to vanish, which is a requirement for a critical point as it enables the divergent responsivity and slow critical dynamics, is that $\text{det }\mathbf{M} = 0$, or~\cite{Woodley2018}
\begin{equation}
(1+a_1b_1)(1+a_2b_2)=a_1a_2c_1c_2.\label{eq:asEurCond}
\end{equation}
As the solution space is now four-dimensional, parametrised e.g.\ by $(p^0_1, p^0_2, \Delta^0_1, \Delta^0_2)$, as opposed to the two-dimensional symmetric space parametrised by $(p,\Delta)$, and the space of critical points is now two- rather than one-dimensional (for example, given any $(p^0_1, p^0_2)$ within some region, there are one or more discrete points $(\Delta^0_1, \Delta^0_2)$ that are critical points), this single condition is not sufficient for a given solution to be a critical point. Rather, it describes a more general three-dimensional space that we shall call the boundary of the unstable region. The intersection of this with the two-dimensional subspace $\tilde{p}_1 = \tilde{p}_2 = 1.75$ (for \mbox{$A_{11}=A_{22}=1$}, \mbox{$A_{12}=A_{21}=2$} as in Section~\ref{sec:symm}) is shown as a thick black line in Fig.~\ref{fig:eur}. In fact Eq.~(\ref{eq:asEurCond}) also encompasses the boundary of the other unstable region that is a generalisation of the points $\text{B}_1$ and $\text{B}_2$; the symmetry-breaking-related one may be specified by $a_{1,2}>0$, i.e.\ both pumps being blue-detuned from their respective Kerr-shifted resonances. We assume for the remainder of this section that Eq.~(\ref{eq:asEurCond}) and $a_{1,2}>0$ hold.
\begin{figure}
\includegraphics*[width=0.95\columnwidth]{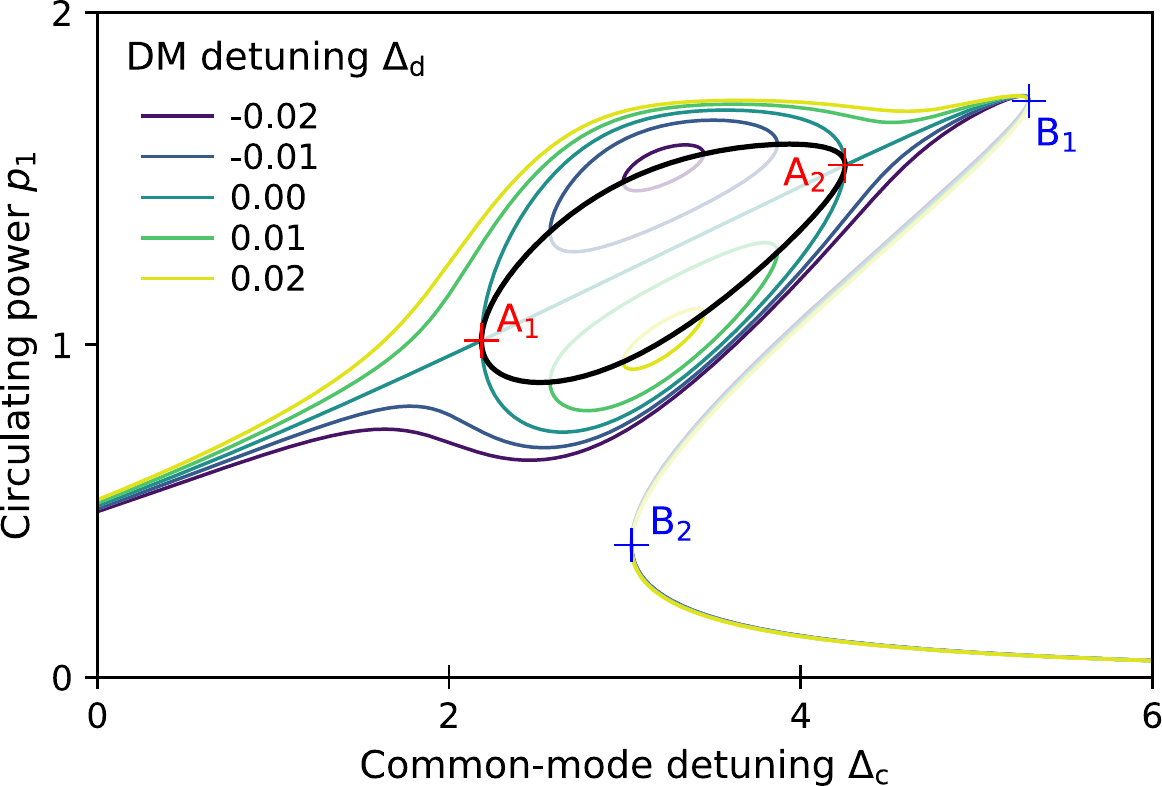}
\caption{Solutions to Eq.~(\ref{eq:ssLorentz}) under symmetric pump powers $\tilde{p}_{1,2}=\tilde{p}=1.75$ as in Fig.~\ref{fig:bubble} but for five values of the differential-mode detuning $\Delta_\text{d}=(\Delta_1-\Delta_2)/2$, plotted as circulating power $p_1$ vs.\ common-mode detuning \mbox{$\Delta_\text{c}=(\Delta_1+\Delta_2)/2$}. The case $\Delta_\text{d}=0$ and points $\text{A}_1$, $\text{A}_2$, $\text{B}_1$, and $\text{B}_2$ are as in Fig.~\ref{fig:bubble}, and faint lines represent unstable solutions. The locus of the edge of the symmetry-breaking-related unstable region for varying $\Delta_\text{d}$ is shown as a thick black line.\label{fig:eur}}
\end{figure}
The next question is how to identify which points on the edge of the unstable region are critical points. To answer this, we proceed in the same way as in Section~\ref{sec:symm}. Defining the quantities
\begin{equation}\begin{aligned}
Q &= \frac{1+a_1b_1}{a_1c_2},\quad R = \frac{1+a_2b_2}{1+a_1b_1},\\
S &= \sqrt{1+a_1b_1+a_2b_2},
\end{aligned}\end{equation}
and noting that (\ref{eq:asEurCond}) implies that
\begin{equation}
Q^2R= \frac{A_{12}\,a_2}{A_{21}\,a_1}\label{eq:asQRrelat},
\end{equation}
the eigenvectors and eigenvalues of $\mathbf{M}$ on the boundary of the unstable region may be written as
\begin{align}
\mathbf{v}_1&=\begin{pmatrix}-QRa_1\\-QR\\a_2\\1\end{pmatrix}\!,\;\lambda_1 = 0;\;\;\mathbf{v}_2=\begin{pmatrix}QRa_1\\-QR\\-a_2\\1\end{pmatrix}\!,\;\lambda_2 = -2;\nonumber\\
\mathbf{v}_3&=\begin{pmatrix}-ia_1Q/S\\Q\\-ia_2/S\\1\end{pmatrix}\!,\;\lambda_3 = -1+iS;\label{eq:asEigens}\\
\mathbf{v}_4&=\begin{pmatrix}ia_1Q/S\\Q\\ia_2/S\\1\end{pmatrix}\!,\;\lambda_4 = -1-iS.\nonumber
\end{align}
The normalisation of the eigenvectors is chosen to be consistent with Eq.~(\ref{eq:eigens}) in the symmetric case. Using these new eigenvalues and eigenvectors and the corresponding inverse basis $\{\mathbf{u}_i\}: \mathbf{u}_i\cdot\mathbf{v}_j = \delta_{ij}$, where again $\delta_{ij}$ is the Kronecker delta, the reasoning in Section~\ref{sec:symm} can be replicated with just two slight modifications. Firstly, whereas in Section~\ref{sec:symm} we have $K_{111}=0$ by symmetry, here that condition specifies the critical points. In other words, it distinguishes the critical points from the rest of the boundary of the unstable region since it implies that, in the absence of external perturbations, $\dot{\mu}_1$ is proportional to $-\mu_1^3$ rather than $\mu_1^2$ to leading order. This is a necessary condition for a critical point since a non-zero $\mu_1^2$ term in $\dot{\mu}_1$ would mean that $\mu_1$ is unstable for one sign of perturbation, somewhat like a particle in an $x^3$ potential. This occurs everywhere on the boundary of the unstable region except for the critical point, where the stability is analogous to a particle in an $x^4$ potential. Conveniently, $K_{112}=-K_{111}$ on the boundary of the unstable region, so $K_{111}=0$ implies $K_{112}=0$, another result of directional symmetry that was used in Section~\ref{sec:symm}. The condition $K_{111}=0$ can be expressed as
\begin{equation}\begin{aligned}
&(1-3a_2^2)(QA_{22}e^0_2-A_{12}e^0_1)\\+\,Q^2R^2&(1-3a_1^2)(QA_{21}e^0_2-A_{11}e^0_1)=0.\label{eq:asCPCond}
\end{aligned}\end{equation}

The second slight change from Section~\ref{sec:symm} is in the resolution of the external perturbations into common- and differential-mode components. In the general asymmetric case, the ratios of the coefficients of $\delta_{1,2}$ and $\epsilon_{1,2}$ are different in every relevant $d_i$ or $D_{ij}$ term. However, it is still true that $d_1$ must scale as $\mu_1^3$ and $d_{3,4}$ and $D_{11}$ as $\mu_1^2$ in order to preserve the natural hierarchy of scalings of terms  in the eigenbasis, and that no other elements of $d_i$ or $D_{ij}$ contribute to leading order. Therefore we may define linear combinations $\delta_\text{c'}$ and $\delta_\text{d'}$ of $\delta_{1,2}$ that scale as $\mu_1^2$ and $\mu_1^3$ respectively, requiring only that $\partial d_1/\partial \delta_\text{c'} = 0$ (and $\partial d_{3,4}/\partial \delta_\text{c'},\, \partial D_{11}/\partial \delta_\text{c'} \ne 0$), and still satisfy the scalings of $d_1$, $d_{3,4}$ and $D_{11}$ to leading order. Since
\begin{equation}
d_1=\frac{Qe^0_2\delta_2 - e^0_1\delta_1}{2Q(1+R)}+\frac{Qa_1(1+a_2^2)e^0_2\epsilon_2 - a_2(1+a_1^2)e^0_1\epsilon_1}{4a_1a_2Q(1+R)}
\end{equation}
we shall do this as follows:
\begin{equation}
\delta_\text{c'} = \frac{1}{2}\left(\delta_1+\frac{Qe^0_2}{e^0_1}\delta_2\right)\!,\quad\delta_\text{d'} = \frac{1}{2}\left(\delta_1-\frac{Qe^0_2}{e^0_1}\delta_2\right)\!.\vspace{2mm} \label{eq:asdeltacpdeltadp}
\end{equation}
Similarly, for $\epsilon_{1,2}$ we define
\begin{equation}\begin{aligned}
\epsilon_\text{c'} &= \frac{1}{2}\left(\epsilon_1+\frac{Qa_1(1+a_2^2)e^0_2}{a_2(1+a_1^2)e^0_1}\epsilon_2\right)\!,\\\epsilon_\text{d'} &= \frac{1}{2}\left(\epsilon_1-\frac{Qa_1(1+a_2^2)e^0_2}{a_2(1+a_1^2)e^0_1}\epsilon_2\right)\!,
\end{aligned}\end{equation}
so that $\partial d_1/\partial \epsilon_\text{c'} = 0$.
Finally, like in Section~\ref{sec:symm}, we can express the dynamics of $\mu_1$ in the form
\begin{equation}
\dot{\mu}_1 = d_1 + D_{11}^\text{eff}\,\mu_1 + L_{1111}^\text{eff}\,\mu_1^3,
\end{equation}
where
\begin{equation}\begin{aligned}
D_{11}^\text{eff} &= D_{11} - 2\left(\frac{K_{113}d_3}{\lambda_3}+\frac{K_{114}d_4}{\lambda_4}\right)\quad\text{and}\\
L_{1111}^\text{eff} &= L_{1111} - 2\left(\frac{K_{113}K_{311}}{\lambda_3}+\frac{K_{114}K_{411}}{\lambda_4}\right)\!,
\end{aligned}\end{equation}
applying the transformation
\begin{equation}
x=D_{11}^\text{eff},\;\; y=-\sqrt{-L_{1111}^\text{eff}}\,\mu_1,\;\; z=-\sqrt{-L_{1111}^\text{eff}}\,d_1
\end{equation}
to reproduce Eq.~(\ref{eq:effCritDynEq}), which works because $L_{1111}^\text{eff}<0$.
The quantities $d_1$, $D_{11}^\text{eff}$ (to leading order) and $L_{1111}^\text{eff}$ (simplified a little by assuming Eq.~(\ref{eq:asCPCond})) are given by:
\begin{widetext}
\begin{equation}
d_1=-\frac{e^0_1(2a_1\delta_\text{d'}+(1+a_1^2)\epsilon_\text{d'})}{2Q(1+R)a_1},
\end{equation}
\begin{equation}\begin{aligned}
D_{11}^\text{eff}=\frac{1}{2a_1^2a_2 Q^2 (1 + R)}\biggl(&\frac{a_1 Q (a_1 (a_2^2-1) e^0_1 + (a_1^2-1)a_2 e^0_2 Q R) \delta_\text{c'}}{e^0_2}\\
 &+\frac{e^0_1}{1 + S^2} \Bigl(a_1 \bigl(Q (a_2 (1-3a_2^2) A_{22} e^0_2 +a_1 (1-a_2^2) A_{21} e^0_1 Q) + 2 A_{12} a_2^2 (a_2 e^0_1 +a_1 e^0_2 Q)\bigr)\\
 &+ a_2 Q \bigl((1-a_1^2) A_{12} a_2 e^0_2 + a_1 (1-3a_1^2) A_{11} e^0_1 Q\bigr) R\Bigr) (2a_1 \delta_\text{c'} + (1 + a_1^2)\epsilon_\text{c'})\biggr),
\end{aligned}\end{equation}
\begin{equation}\begin{aligned}
L_{1111}^\text{eff}= \frac{1}{2a_1^2a_2(1 + R)}\biggl(&a_1\bigl(a_1(1-a_2^4)A_{22} + 2A_{12}a_2(1-a_1^2a_2^2)R + (1-a_1^4) A_{11}a_2Q^2R^3\bigr)\\
&- \frac{2}{Q^2 (1 + S^2)}\bigl(2 a_1 A_{12} (1 - a_2^2) e^0_1 - a_1 (1 - 3 a_2^2) A_{22} e^0_2 Q + (1 + a_1^2) A_{12} a_2 e^0_2 Q R\bigr)\\
&\bigl(a_1 (Q (a_2 (1 - 3 a_2^2) A_{22} e^0_2 + a_1 (1 - a_2^2) A_{21} e^0_1 Q) + 2 A_{12} a_2^2 (a_2 e^0_1 + a_1 e^0_2 Q))\\
&+ a_2 Q ((1 - a_1^2) A_{12} a_2 e^0_2 + a_1 (1 - 3 a_1^2) A_{11} e^0_1 Q) R\bigr)\biggr).
\end{aligned}\end{equation}
\end{widetext}
The deviations $\delta p_{1,2}$ of the circulating powers $p_{1,2}$ from their steady-state values $p^0_{1,2}$ are given to leading order by $2e^0_if_i$, or
\begin{equation}\begin{aligned}
\delta p_1 = -2QRa_1e^0_1\mu_1,\quad\delta p_2 = 2a_2e^0_2\mu_1.\label{eq:asCircPowDevs}
\end{aligned}\end{equation}
As a final comment, it is worth noting that microresonators generally possess strong thermal nonlinearity due to a combination of thermorefractive effects and thermal expansion~\cite{Carmon2004}. This typically creates circulating-power-dependent resonance frequency shifts between one and two orders of magnitude larger than the Kerr shifts, but which require much longer timescales to take effect, and could thus greatly complicate the critical dynamics. Importantly, however, these effects depend only on the total circulating power $p_1+p_2$ and create equal shifts for both directions, i.e.~change $\Delta_1$ and $\Delta_2$, or equivalently $\delta_1$ and $\delta_2$, by the same amount, assuming that the two modes occupy the exact same region. This means that they can be decoupled from the critical dynamics in two ways -- firstly by making $\delta p_1$ and $\delta p_2$ in Eq.~(\ref{eq:asCircPowDevs}) equal and opposite, and secondly by making the coefficients of $\delta_1$ and $\delta_2$ in $\delta_\text{d'}$ (Eq.~(\ref{eq:asdeltacpdeltadp})) equal and opposite. The latter condition may be written as
\begin{equation}
Qe^0_2=e^0_1,
\end{equation}
whilst the former simplifies via Eq.~(\ref{eq:asQRrelat}) to
\begin{equation}
QA_{21}e^0_2=A_{12}e^0_1.
\end{equation}
Thus the two conditions are equivalent if $A_{21}=A_{12}$, which is in fact necessarily true due to the reciprocity of the Kerr effect, and since the normalisation factor for the circulating power is the same for both modes.
\section{Conclusion and Outlook}
We have derived a theory that explains the dynamics of a bidirectionally-pumped optical resonator with Kerr nonlinearity in the region close to the critical point of the symmetry breaking between counterpropagating light. This was done first for the case of a perfectly symmetrical system in Section~\ref{sec:symm}, before being generalised to asymmetrical pumping conditions and SPM and XPM coefficients in Section~\ref{sec:asymm}. A condition for compensating the various asymmetries with each other to recover a critical point is derived (Eqs.~(\ref{eq:asEurCond}) and~(\ref{eq:asCPCond})). The critical dynamics are shown to be described by the simple Eq.~(\ref{eq:effCritDynEq}) in both the symmetric and asymmetric cases, for each of which explicit formulae for the conversion factors to the generalised variables $x$, $y$ and $z$ are obtained. From Eq.~(\ref{eq:effCritDynEq}), we see that the system exhibits scaling invariance, divergent steady-state responsivity and critical slowing down, all of which are universal features of critical systems. Finally, a condition for decoupling the critical dynamics from thermal nonlinearities is discussed.

The theory presented here describes in detail the response of critical-point-enhanced sensors such as gyroscopes~\cite{Kaplan1981, Wang2014a} and refractive index sensors~\cite{Wang2015}. Furthermore, it is applicable to any optical resonator in which two modes interact via the Kerr nonlinearity, including modes of different frequencies, propagation angles~\cite{Haelterman1991} or opposite circular polarisations~\cite{areshev1983polarization, haelterman1994polarization, Copie2019, garbin2019asymmetric}. It also extends to other Kerr-like effects such as the magnetic nonlinearity~\cite{Martin2010}, and even to similar nonlinear systems outside the optical domain.

\begin{acknowledgments}
The authors would like to thank Lewis Hill, Michael Woodley and Leonardo Del Bino for helpful discussions. This work was supported by the Royal Academy of Engineering and the Office of the Chief Science Adviser for National Security under the UK Intelligence Community Postdoctoral Fellowship Programme. The authors also acknowledge funding from H2020 Marie Sklodowska-Curie Actions (MSCA) (No. 748519, CoLiDR), H2020 European Research Council (ERC) (No. 756966, CounterLight), and National Physical Laboratory Strategic Research.
\end{acknowledgments}

\bibliography{TheoryPaper_v2}
\end{document}